\documentclass[copyright]{eptcs}
\providecommand{\event}{QPL 2012}
\usepackage{amssymb,amsmath,stmaryrd,mathrsfs}
\usepackage{mathtools}
\usepackage{xspace}             % put spaces after a \command in text
\usepackage{ifmtarg}
\usepackage{braket}
\usepackage[all]{xy}
\usepackage{url}
\usepackage{enumitem}

\title{Quantum Gauge Field Theory \\ in  \\ Cohesive Homotopy Type Theory}
\author{
  Urs Schreiber
  \institute{University Nijmegen\\ The Netherlands}
  \email{urs.schreiber@gmail.com}
  \and
  Michael Shulman
  \institute{University of San Diego\\
    San Diego, CA, USA}
  \email{shulman@sandiego.edu}}
\def\titlerunning{QGFT in CoHoTT}
\def\authorrunning{U. Schreiber \& M. Shulman}

\newcommand{\type}{\ensuremath{\mathsf{Type}}\xspace}
\newcommand{\ig}{\ensuremath{\infty\text{-}\mathbf{Gpd}}\xspace}
\renewcommand{\H}{\ensuremath{\mathbf{H}}\xspace}
\newcommand{\io}{\ensuremath{(\infty,1)}}
\newcommand{\sharpsub}[1]{\sharp_{\scriptscriptstyle #1}}
\newcommand{\E}{\ensuremath{\mathcal{E}}\xspace}
\newcommand{\M}{\ensuremath{\mathcal{M}}\xspace}
\newcommand{\esc}{\ensuremath{\mathord{\uparrow}}}

\newcommand{\factsharp}{\textsf{fact}_{\sharp}}
\newcommand{\factflat}{\textsf{fact}_{\flat}}

\newcommand{\ssl}{\mathord{\sslash}}

\makeatletter
\def\jd#1{\@jd#1\ej}
\def\@jd#1|-#2\ej{\@@jd#1,,\;\vdash\;\left(#2\right)}
\def\@@jd#1,{\@ifmtarg{#1}{\let\next=\relax}{\left(#1\right)\let\next=\@@@jd}\next}
\def\@@@jd#1,{\@ifmtarg{#1}{\let\next=\relax}{,\,\left(#1\right)\let\next=\@@@jd}\next}
\def\jdm#1{\@jdm#1\ej}
\def\@jdm#1|-#2\ej{\@@jd#1,,\\\vdash\;\left(#2\right)}
\def\cm{,}
\makeatother

\begin{document}

\maketitle

\begin{abstract}
  We implement in the formal language of \emph{homotopy type theory} a new set of axioms called \emph{cohesion}.
  Then we indicate how the resulting \emph{cohesive homotopy type theory} naturally serves
  as a formal foundation for central concepts in quantum gauge field theory.
  This is a brief survey of work by the authors developed in detail elsewhere~\cite{ShulmanCohesion,SchreiberCohesion}.
\end{abstract}

%\tableofcontents

\section{Introduction}
\label{sec:introduction}

The observable world of physical phenomena is fundamentally governed by
\emph{quantum gauge field theory} (QFT) \cite{Deligne},
as was recently once more confirmed by the
detection \cite{CERNPressRelease} of the Higgs boson \cite{Bernstein}.
%edit
On the other hand, the world of mathematical concepts
is fundamentally governed by \emph{formal logic}, as elaborated in a foundational system such as axiomatic set theory or type theory.

Quantum gauge field theory is traditionally valued for the elegance and beauty
of its mathematical description (as far as this has been understood).
Formal logic is likewise valued for elegance and simplicity; aspects which have become especially important recently because they enable formalized mathematics to be verified by computers.
This is generally most convenient using type-theoretic foundations~\cite{MartinLoef}; see e.g.~\cite{Coquand}.

However, the mathematical machinery of quantum gauge field theory ---
such as differential geometry
(for the description of spacetime \cite{GS}),
differential cohomology
(for the description of gauge force fields \cite{HopkinsSinger, Freed})
and symplectic geometry
(for the description of geometric quantization \cite{Brylinski})
--- has always seemed to be many levels of complexity above the mathematical foundations.
Thus, while automated proof-checkers
can deal with fields like linear algebra \cite{Coquand},
even formalizing a basic differential-geometric \emph{definition} (such as a principal connection on a smooth manifold) seems intractable,
not to speak of proving its basic properties.

\medskip

We claim that this situation improves drastically by combining two insights from type theory.
The first is that type theory can be interpreted ``internally''
in
%edit:
%a wide variety of categories
locally Cartesian closed categories
(see e.g.~\cite{CD}),
including categories of smooth spaces (which contain the classical category of smooth manifolds).
In this way, type-theoretic arguments which appear to speak about discrete sets may be interpreted to speak about smooth spaces, with the smooth structure automatically ``carried along for the ride''.
Thus, differential notions can be developed in a simple and elegant axiomatic framework --- avoiding the complexity of the classical definitions by working in a formal system whose \emph{basic objects} are already ``smooth''.
%If one additionally posits certain ``infinitesimal objects'',
%one arrives at
This is known as \emph{synthetic differential geometry} (SDG)~\cite{kock:sdg,lavendhomme:sdg,SIA}.
In this paper, we
% will take a different more general
will axiomatize it in a way which does not require that our basic objects are ``smooth'',
only that they are ``cohesive'' as in~\cite{Lawvere}.
This includes topological objects as well as smooth ones,
and also variants of differential geometry such as \emph{supergeometry},
which is necessary for a full treatment of quantum field theory (for the description of fermions).

The second
insight is that types in
%edit
intensional
type theory can also behave like \emph{homotopy types},
a.k.a.\ \emph{$\infty$-groupoids}, which are not just sets of points but contain higher homotopy information.
Just as SDG imports differential geometry directly into type theory, this one imports \emph{homotopy theory}, and as such is called \emph{homotopy type theory}~\cite{Voevodsky,AwodeyHoTT,ShulmanCourse,hottbook}.

Combining these insights, we obtain \emph{cohesive homotopy type theory}~\cite{ShulmanCohesion,SchreiberCohesion}.
Its basic objects (the ``types'') have both \emph{cohesive structure} and \emph{higher homotopy structure}.
%We emphasize that
These two kinds of structure are independent, in contrast to how classical algebraic topology identifies homotopy types with the topological spaces that present them.
For instance, the geometric circle $S^1$ is categorically \emph{0-truncated}
(it has a mere set of points with no isotropy),
but carries an interesting topological or smooth structure ---
whereas the homotopy type it presents, denoted $\Pi(S^1)$ or $\mathbf{B}\mathbb{Z}$
(see below), has (up to equivalence) only \emph{one} point (with trivial topology), but that point has a countably infinite isotropy group.
More general cohesive homotopy types can be nontrivial in both ways,
such as orbifolds~\cite{MoerdijkPronk,Lerman} and moduli stacks~\cite{Wang}.
%edit:
In particular the standard model of SDG known as the {\it Cahiers topos} lifts to a model of
cohesive homotopy type theory that contains orbifolds and generally moduli stacks equipped with
synthetic differential structure (section 4.5 of \cite{SchreiberCohesion}).

\medskip

Today it is clear that homotopy theory is at the heart of quantum field theory.
One way to define an
$n$-dimensional QFT is as a rule that assigns to each
closed $(n-1)$-dimensional manifold, a vector space --- its space of \emph{quantum states} --- and
to each $n$-dimensional \emph{cobordism}, % between manifolds
a linear map between the corresponding vector spaces --- a \emph{correlator} ---
in a way which respects gluing of cobordisms.
An \emph{extended} or \emph{local} QFT also assigns data
to all $(0 \leq k \leq n)$-dimensional manifolds,
such that the data assigned to any manifold can be reconstructed by gluing along lower dimensional boundaries.
By~\cite{BD,LurieTQFT},
the case of \emph{extended topological QFT}
(where the manifolds are equipped only with smooth structure) is entirely defined and classified by
a universal construction in \emph{directed homotopy theory}, i.e.\ $(\infty,n)$-category theory~\cite{BarwickSchommer-Pries,ShulmanDirected}.
For non-topological QFTs, where the cobordisms have conformal or metric
structure, the situation is more complicated, but directed
homtopy type theory still governs the construction;
see~\cite{SatiSchreiber} for recent developments.
%new paragraph
We will see in section \ref{GeometricQuantization}
how in cohesive homotopy type theory one naturally obtains such higher
spaces of (pre-)quantum states assigned to manifolds of higher codimension.

%new paragraph
Notice, as discussed in the introduction of \cite{SatiSchreiber},
that many of the developments that led to these insights
about the role of homotopy theory in quantum field theory originate in \emph{string theory},
but they are now statements in pure QFT. String theory is the investigation
of the \emph{second quantization} of 1-dimensional quanta (``strings'') instead of 0-dimensional
quanta (``particles'') and this increase in geometric dimension induces a corresponding
increase of homotopical dimension of many aspects. At least some of these are also usefully captured
by cohesive homotopy type theory \cite{SchreiberLectures}.
But while the phenomenological relevance of string theory for fundamental physics remains speculative,
its proposal alone led to further investigation into the nature of pure quantum field theories
which showed how this alone already involves homotopy theory, as used here.

The value of \emph{cohesive} homotopy theory for physics lies in the observation that
the QFTs observed to
govern our world at the fundamental level --- namely,
Yang-Mills theory (for electromagnetism and the weak and strong nuclear forces)
and Einstein general relativity (for gravity) --- are not random
instances of such QFTs.
Instead
(1) their construction follows a geometric principle,
traditionally called the \emph{gauge symmetry principle}~\cite{RS},
and (2) they are obtained by \emph{quantization} from
(``classical'') data that lives in \emph{differential cohomology}.
Both of these aspects are actually native to cohesive homotopy type theory, as follows:
% It is noteworthy that both of these aspects are secretly native to cohesive homotopy
% (type) theory:
\begin{enumerate}[leftmargin=*]
\item The very concept of a collection of quantum field configurations
with \emph{gauge transformations} between them
is really that of a configuration \emph{groupoid}, hence of a homotopy 1-type.
More generally, in higher gauge theories such as those that appear in string theory, the
\emph{higher gauge symmetries} make the configurations form higher groupoids, hence general
homotopy types.
Furthermore,  these configuration groupoids of gauge fields have
\emph{smooth structure}: they are \emph{smooth homotopy types}.

To physicists, these smooth configuration groupoids
are most familiar in their infinitesimal Lie-theoretic
approximation: the (higher) \emph{Lie algebroids} whose function algebras
constitute the \emph{BRST complex}, in terms of which modern quantum gauge theory is
formulated \cite{HT}. The degree-$n$
\emph{BRST cohomology} of these complexes
corresponds to the $n^{\mathrm{th}}$ \emph{homotopy group} of the cohesive homotopy types.
\item Gauge fields are \emph{cocycles} in a cohomology theory
(sheaf hyper-cohomology), and the
%(higher)
gauge transformations
between them are its %(higher)
coboundaries.
% in such a cohomology theory.
%But
A classical result~\cite{Brown} says, in modern language,
that all such cohomology theories are realized in some interpretation of
homotopy type theory (in some $(\infty,1)$-category of $(\infty,1)$-sheaves):
the cocycles on $X$ with coefficients in $A$ are just functions $X \to A$.
%and the space of such functions is the
%smooth homotopy type of cocycles and higher coboundaries:
Moreover, if $A$ is the coefficients of some \emph{differential} cohomology theory, then the \emph{type} of all such functions is exactly the configuration gauge groupoid of quantum fields on $A$ from above.
% More specifically, gauge fields are cocycles for $A$ the coefficient of
%a \emph{differential} cohomology theory.
This cannot be expressed in plain homotopy theory, but it can be in \emph{cohesive} homotopy theory.

Finally, differential cohomology is also the natural context for
\emph{geometric quantization}, so that central aspects of this process can also be formalized in
cohesive homotopy type theory.
\end{enumerate}
%Therefore, cohesive homotopy type theory provides a very natural language in which to formalize such quantum field theories \emph{directly}.

\vspace{.2cm}

In \S\ref{sec:cohott}, we briefly review homotopy type theory and then describe the axiomatic formulation of cohesion.
The axiomatization is chosen so that if we \emph{do} build things from the ground up out of sets, then we can construct categories (technically, \io-categories of \io-sheaves) in which cohesive homotopy type theory is valid internally.
% This involves various aspects of interest in their own right, such as the internal formalization of reflective sub-\io-categories.
This shows that the results we obtain can always be referred back to a classical context.
However, we emphasize that the axiomatization stands on its own as a formal system.

Then in \S\ref{sec:quantum}, we show how cohesive homotopy type theory directly expresses
fundamental concepts in differential geometry, such as differential forms,
Maurer-Cartan forms, and connections on principal bundles.
Moreover, by the homotopy-theoretic ambient logic, these concepts are thereby automatically generalized to
\emph{higher differential geometry} \cite{NSS}. In particular,
we show how to naturally formulate
\emph{higher moduli stacks} for \emph{cocycles in differential cohomology}.
Their 0-truncation shadow has been known to
formalize gauge fields and higher gauge fields \cite{Freed}.
We observe that their full homotopy formalization yields a refinement of
the Chern-Weil homomorphism from secondary characteristic classes to
cocycles, and also the action functional of generalized
Chern-Simons-type gauge theories with an \emph{extended} geometric prequantization.
At present, however, completing the process of quantization requires special properties of the usual models; work is in progress isolating exactly how much quantization can be done formally in cohesive homotopy type theory.

The constructions of \S\ref{sec:cohott} have been fully implemented in {\tt Coq}%
%edit
\footnote{Currently, this requires a patch to Coq that collapses universe levels (making it technically inconsistent).  However, true ``universe polymorphism'' is slated for future inclusion in Coq, which should should make this no longer necessary.} \cite{Coq};
the source code can be found at~\cite{CohesiveCoq}.  With this as foundation, the implementation
of much of \S\ref{sec:quantum} is straightforward.

\section{Cohesive homotopy type theory}
\label{sec:cohott}

\subsection{Categorical type theory}
\label{sec:type-theory}

A type theory is a formal system
%, which behaves in many respects like a programming language, but can also be used as a foundation for mathematics.
whose basic objects are \emph{types} and \emph{terms},
and whose basic assertions are that a term $a$ belongs to a type $A$, written ``$a:A$''.
% As a first approximation (which we will improve in a moment), types can be thought of as like \emph{sets} and terms as like \emph{elements} of these sets.
% Unlike in traditional set theory, however, a term can belong to only one type (at least in the type theories we are concerned with).
% This matches the actual practice of mathematics more closely than does the global membership relation in traditional set theory.
%
%One writes $a:A$ to mean that the term $a$ belongs to the type $A$.
%Thus, for instance, ``$\sqrt{2}:\mathbb{R}$'' indicates that the term $\sqrt{2}$ belongs to the type $\mathbb{R}$ of real numbers.
%Terms can also contain variables, each of which must be declared to have a particular type.
More generally, $\jd{x:A, y:B |- c:C}$ means that given variables $x$ and $y$ of types $A$ and $B$, the term $c$ has type $C$.
%(Type theorists generally omit the parentheses, but we find them helpful, especially while learning to parse the notation.)
%For instance, $\jd{x:\mathbb{R}, y:\mathbb{R} |- x+y:\mathbb{R}}$ indicates that given two real numbers $x,y$, their sum $x+y$ is again a real number.
%
For us,
% will always be \emph{dependent type theories with universes}, which means that
types themselves are terms of type \type.
(One avoids paradoxes arising from $\type:\type$ with a hierarchy of universes.)
%
% Since \type is itself a type, one might expect that $\type:\type$.
% But unsurprisingly, this leads to the same sort of contradictions as does a set of all sets in traditional set theory.
% One resolves this by introducing a hierarchy of types with $\type_0 : \type_1$ and $\type_1 : \type_2$, etc.
% We will say no more about this.
%
%edit
Types involving variables are called \emph{dependent types}.

Operations on types include cartesian product $A\times B$, disjoint union $A+B$, and function space $A\to B$, each with corresponding rules for terms.
Thus $A\times B$ contains pairs $(a,b)$ with $a:A$ and $b:B$, while $A\to B$ contains functions $\lambda x^A.b$ where $b:B$ may involve the variable $x:A$, i.e.\ $\jd{x:A |- b:B}$.
%
% Thus, for instance, $\jd{X:\type, Y:\type |- X\times Y:\type}$ indicates that given two types $X$ and $Y$, their cartesian product $X\times Y$ is again a type.
% Other operations on types include the disjoint union, written $X+Y$, and the type of functions from $X$ to $Y$, which type theorists usually write as $X\to Y$.
% (Thus, the notation $f:X\to Y$ for a term belonging to the type $X\to Y$ matches the traditional notation asserting that $f$ is a function from $X$ to $Y$.)
%
% For instance, $\jd{x:\mathbb{R} |- \Set{y:\mathbb{R} | y<x}:\type}$ indicates that for any real number $x$, the collection of real numbers less than $x$ is a type; together these types form a type \emph{dependent on} $x$.
% There are also operations on dependent types.
Similarly, if $\jd{x:A |- B(x):\type}$ is a dependent type, its \emph{dependent sum} $\sum_{x:A} B(x)$ contains pairs $(a,b)$ with $a:A$ and $b:B(a)$, while its \emph{dependent product} $\prod_{x:A} B(x)$ contains functions $\lambda x^A .b$ where $\jd{x:A |- b : B(x)}$.

% Hopefully it is unsurprising that such a system, together with appropriate rules for constructing types and terms, can serve as a foundation for mathematics in much the same way as traditional set theory does.
In many ways, types and terms behave like sets and elements as a foundation for mathematics.
One fundamental difference is that
% (which takes some getting used to)
in type theory, rather than proving theorems \emph{about} types and terms, one \emph{identifies}
``propositions'' with
% theorems with particular types, and proofs with particular terms.
% Specifically, a \emph{proposition} is a
types containing at most one term
%edit:
(also called ``mere propositions'', for emphasis),
% (up to a suitable notion of equivalence; we will come back to this below),
and ``proofs'' with terms belonging to such types.
%edit:
Constructions such as $\times,\to,\prod$ restrict to logical operations such as $\wedge,\Rightarrow,\forall$% on propositions
, embedding logic into type theory.
By default, this logic is \emph{constructive}, but one can force it to be classical.

Type theory also admits \emph{categorical models}, where types are interpreted by objects of a category $\H$, while a term $\jd{x:A, y:B |- c:C}$ is interpreted by a morphism $A\times B\to C$.
A dependent type $\jd{x:A |- B(x):\type}$ is interpreted by $B\in\H/A$, while $\jd{x:A|- b:B(x)}$ is interpreted  by a \emph{section} of $B$, and substitution of a term for $x$ in $B(x)$ is interpreted by pullback.
Thus, we can ``do mathematics'' internal to \H, with any additional structure on its objects
%type theory is a set-like syntax for objects and morphisms in abstract categories, where additional structure
%(such as cohesion or higher homotopies)
carried along automatically.
In this case, the logic is usually unavoidably constructive.

In the context of quantum physics, such ``internalization'' has been used in the ``Bohrification'' program~\cite{Bohrification} to make noncommutative von Neumann algebras into internal commutative ones.
%  one constructs a category containing an internal \emph{commutative} von Neumann algebra.
There are also ``linear'' type theories which describe mathematics internal to monoidal categories (such as Hilbert spaces); see~\cite{Rosetta}.

\subsection{Homotopy type theory}
\label{sec:hott}

Since propositions are types, we expect \emph{equality types}
\[\jd{x:A,y:A|- (x=y):\type}.\]
But surprisingly, $(x=y)$ is naturally \emph{not} a mere proposition. % (i.e.\ they can contain more than one element).
We can add axioms forcing it to be so, but if we don't, we obtain \emph{homotopy type theory}~\cite{hottbook}, where types behave less like sets and more like \emph{homotopy types} or \emph{$\infty$-groupoids}.\footnote{The associativity of terminology ``homotopy (type theory) = (homotopy type) theory'' is coincidental, though fortunate!}
Space does not permit an introduction to homotopy theory and higher category theory here; see e.g.~\cite[\S1.1]{LurieHTT}.
We re-emphasize that in \emph{cohesive} homotopy type theory, simplicial or algebraic models for homotopy types are usually less confusing than topological ones.

Homotopy type theory admits models in \emph{\io-categories}, where the equality type of $A$ indicates its diagonal $A\to A\times A$.
%Roughly, nontrivial homotopy theory arises because such diagonals need not be monic.
%
% , which are to categories in the same the way that $\infty$-groupoids are to sets.
% In addition to objects and morphisms, an \io-category has 2-equivalences between morphisms, 3-equivalences between 2-equivalences, etc., and we generally only ever ask whether morphisms are equivalent rather than equal.
% All the rules of interpreting type theory are essentially the same; the difference is that the diagonals $B\to B\times B$ need not be monomorphisms.
% In particular, now to give a term belonging to the dependent type $\jd{x:A |- (f(x)=g(x)):\type}$ means to give an \emph{equivalence} between the morphisms $f$ and $g$.
%
Voevodsky's \emph{univalence axiom}~\cite{KLVunivalence} implies that the type \type is an
%This also allows us to give a good interpretation of the type \type.
%It is an
\emph{object classifier}: there is a morphism $p:{\type}_\bullet\to \type$ such that
%every $B\to A$ is the pullback of $p$ along an essentially unique morphism $A\to \type$.
%More precisely,
pullback of $p$ induces an equivalence of $\infty$-groupoids
\begin{equation}
  \H(A,\type) \simeq \mathrm{Core}_\kappa(\H/A).\label{eq:objclassif}
\end{equation}
(The \emph{core} of an \io-category contains all objects, the morphisms that are equivalences, and all higher cells; $\mathrm{Core}_\kappa$ denotes the
%edit
small full subcategory of ``$\kappa$-small'' objects.)
% Having object classifiers in this sense requires $\H$ to be an \io-category
%: for instance if $\H$ is a mere category, then $\mathrm{Core}(\H/A)$ would be a groupoid, but $\H(A,\type)$ would be only a set, so we could not expect them to be equivalent.
%
%Actually, to avoid paradoxes, we only demand~\eqref{eq:objclassif} for some small sub-$\infty$-groupoid of $\mathrm{Core}(\H/A)$.
%Thus, there can be multiple object classifiers which classify objects of different sizes.
%
A well-behaved \io-category with object classifiers is called an \emph{\io-topos}; these are the natural places to internalize homotopy type theory.\footnote{There are, however, coherence issues in making this precise, which are a subject of current research; see e.g.~\cite{ShulmanUnivalence,ShulmanUnivalence2}.}
%However, in the worst case it should be possible to slightly modify the type theory so as to make it true, so we will not bother about it.)

An object $X\in\H$ is \emph{$n$-truncated} if it has no homotopy above level $n$.
The $0$-truncated objects are like sets, with no higher homotopy, while the $(-1)$-truncated objects are the propositions.
The $n$-truncated objects in an \io-topos are reflective, with reflector $\pi_n$; in type theory, this is
%edit
an example of
a \emph{higher inductive type}~\cite{ShulmanLumsdaine,hottbook},%ShulmanHITs,LumsdaineHITs,
%edit
which is the homotopy-theoretic refinement of a type defined by induction, such as the natural numbers.
The $(-1)$-truncation of a morphism $X\to Y$, regarded as an object of $\H/Y$, is its \emph{image factorization}.

\subsection{Cohesive \io-toposes}
\label{sec:cohesion}

A \emph{cohesive \io-topos} is an \io-category whose objects can be thought of as $\infty$-groupoids
%In our applications, $\H$ will be a \emph{cohesive \io-topos}.
%This means we can think of its objects as $\infty$-groupoids whose elements and higher cells are additionally
endowed with ``cohesive structure'', such as a topology or a smooth structure.
As observed in~\cite{Lawvere} for 1-categories, this gives rise to a string of adjoint functors relating $\H$ to \ig (which replaces $\mathbf{Set}$ in~\cite{Lawvere}).
First, the \emph{underlying} functor $\Gamma:\H\to\ig$ forgets the cohesion.
This can be identified with the hom-functor $\H(*,-)$, where the terminal object $*$ is a single point with its trivial cohesion.

Secondly, any $\infty$-groupoid admits both a \emph{discrete} cohesion, where no distinct points ``cohere'' nontrivially, and a \emph{codiscrete} cohesion, where all points ``cohere'' in every possible way.
This gives two fully faithful functors $\Delta: \ig\to\H$ and $\nabla:\ig\to\H$, left and right adjoint to $\Gamma$ respectively.
%(In thinking about the adjoint triple $\Delta\dashv\Gamma\dashv\nabla$, we encourage the reader to think about the like-named functors between the categories of topological spaces and sets.)

Finally, $\Delta$ also has a left adjoint $\Pi$, which preserves finite products.
In~\cite{Lawvere}, $\Pi$ computes \emph{sets of connected components}, but for \io-categories, $\Pi$ computes entire \emph{fundamental $\infty$-groupoids}.
There are two origins of higher morphisms in $\Pi(X)$: the higher morphisms of $X$, and the cohesion of $X$.
If $X=\Delta Y$ is an ordinary $\infty$-groupoid with discrete cohesion, then $\Pi(X) \simeq Y$.
%(this follows formally from the facts that $\Pi\dashv \Delta$ and $\Delta$ is fully faithful).
But if $X$ is a plain set with some cohesion (such as an ordinary smooth manifold), then $\Pi(X)$ is its ordinary fundamental $\infty$-groupoid, whose higher cells are paths and homotopies in $X$.
If $X$ has both higher morphisms and cohesion, then $\Pi(X)$ automatically combines these two sorts of higher morphisms sensibly, like the Borel construction of an orbifold.

Thus, we define an \io-topos \H to be \emph{cohesive} if it has an adjoint string
\begin{equation}
\vcenter{\xymatrix{
  \H \ar@<2mm>[d]|{\Gamma} \ar@<-6mm>[d]|{\Pi}\\
  \ig \ar@<2mm>[u]|{\Delta} \ar@<-6mm>[u]|{\nabla}
}}\label{eq:cohesion}
\end{equation}
where $\Delta$ and $\nabla$ are fully faithful and $\Pi$ preserves finite products.
Using $\infty$-sheaves on sites~\cite{CohesiveSites}, we can obtain such \H's which contain smooth manifolds as a full subcategory; we call these \emph{smooth models}.

Now we plan to work in the internal type theory of such an \H, so we must reformulate cohesiveness internally to \H.
But since $\Delta$ and $\nabla$ are fully faithful, from inside \H we see two subcategories, of which the codiscrete objects are reflective, and the discrete objects are both reflective (with reflector preserving finite products) and coreflective.
We write $\sharp \coloneqq \nabla\Gamma$ for the codiscrete reflector, $\flat \coloneqq \Delta\Gamma$ for the discrete coreflector, and $\Pi \coloneqq \Delta\Pi$ for the discrete reflector.
Assuming only this, if $A$ is discrete and $B$ is codiscrete, we have
\[ \H(\sharp A,B) \simeq \H(A,B) \simeq \H(A,\flat B) \]
so that $\sharp\dashv\flat$ is an adjunction between the discrete and codiscrete objects.
If we assume that for \emph{any} $A\in\H$, the maps $\flat A\to \flat\sharp A$ and $\sharp \flat A\to\sharp A$ induced by $\flat A\to A\to \sharp A$ are equivalences, this adjunction becomes an equivalence, modulo which $\flat$ is identified with $\sharp$ (i.e.\ $\Gamma$).
From this we can reconstruct~\eqref{eq:cohesion}, except that the lower \io-topos need not be \ig.
This is expected: just as homotopy type theory admits models in all \io-toposes, cohesive homotopy type theory admits models that are ``cohesive over any base''.

We think of $\sharp$, $\flat$, and $\Pi$ as \emph{modalities}, like those of~\cite{AB}, but which apply to all types, not just propositions.
Note also that as functors $\H\to \H$, we have $\Pi\dashv\flat$, since for any $A$ and $B$
\[ \H(\Pi A,B)\simeq \H(\Pi A,\flat B) \simeq \H(A,\flat B) \]
as both $\Pi A$ and $\flat B$ are discrete.

\subsection{Axiomatic cohesion I: Reflective subfibrations}
\label{sec:axiomatic-cohesion-i}

% Now we want to describe the structure of a cohesive \io-topos \H purely in terms of its internal type theory.
% The easiest part of this is the adjunction $\Gamma \dashv \nabla$.
% Since $\nabla$ is fully faithful, this adjunction is determined by a reflective subcategory of \H (the \emph{codiscrete objects}) which is equivalent to \ig.
% We will not be able to say, internally to \H, that this reflective subcategory is equivalent to \ig, but it turns out that this doesn't matter.
% In fact, it makes the theory more general, able to describe cohesion ``over arbitrary bases''.

We begin our internal axiomatization with the reflective subcategory of codiscrete objects.
An obvious
way to describe a reflective subcategory in type theory is to use \type.
%edit
(This idea is also somewhat na\"ive; in a moment we will discuss an important subtlety.)
First we need, for any type $A$, a proposition expressing the assertion ``$A$ is codiscrete''.
Since propositions are particular types, this can simply be a function term
\begin{equation}
  \mathsf{isCodisc} : \type \to \type
\end{equation}
together with an axiom asserting that for any type $A$, the type $\mathsf{isCodisc}(A)$ is a proposition:
\begin{equation}
  \jd{A:\type |- \mathsf{isCodiscIsProp}_A : \mathsf{isProp} (\mathsf{isCodisc}(A))}
\end{equation}
Next we need the reflector $\sharp$ and its unit:
\begin{gather}
\sharp : \type\to\type.\\
%We have an axiom saying that $\sharp A$ is always codiscrete:
\jd{A:\type |- \mathsf{sharpIsCodisc}_A : \mathsf{isCodisc}(\sharp A)}\\
%Every type has a map to its codiscrete reflection (the unit of the adjunction $\Gamma \dashv \nabla$):
  \jd{A:\type |- \eta_A : A \to \sharp A}.  \label{SharpUnit}
\end{gather}
Finally, we assert the universal property of the reflection: if $B$ is codiscrete, then the space of morphisms $\sharp A \to B$ is equivalent, by precomposition with $\eta_A$, to the space of morphisms $A\to B$.
\begin{equation}
  \jd{A:\type, B:\type, \mathsf{bc} : \mathsf{isCodisc}(B) |-
    \mathsf{tsr} : \mathsf{isEquiv} (\lambda f^{\sharp A\to B}. f\circ \eta_A)
  }\label{eq:tsr}
\end{equation}
This looks like a complete axiomatization of a reflective subcategory, but in fact it describes more data than we want, because \type is an object classifer for \emph{all} slice categories.
If \H satisfies these axioms, then each $\H/X$ is equipped with a reflective subcategory, and moreover these subcategories and their reflectors are stable under pullback.
For instance, if $A\in\H/X$ is represented by $\jd{x:X |- A(x):\type}$, then the dependent type $\jd{x:X |- \sharp(A(x)):\type}$ represents a ``fiberwise reflection'' $\sharpsub{X}(A) \in \H/X$.
%Of course, $\sharpsub{X}(A)$ will generally be different from the global reflection $\sharp A$, which need not even admit a map to $X$.

We call such data a \emph{reflective subfibration}~\cite{CJKP}.
%; its existence also has nontrivial implications for the underlying reflective subcategory.
If a reflective subcategory underlies some reflective subfibration, then its reflector preserves finite products, and the converse holds in good situations~\cite{ShulmanSubfibrations}.
%This is, of course, the case for $\sharp = \nabla \Gamma$: since both $\nabla$ and $\Gamma$ are right adjoints, $\sharp$ even preserves all finite \emph{limits}.
If the reflector even preserves all finite \emph{limits}, as $\sharp$ does, then there is a \emph{canonical} extension to a reflective subfibration.
Namely, we define $A\in\H/X$ to be \emph{relatively codiscrete} if the naturality square for $\eta$:
\[\vcenter{\xymatrix@-.5pc{
    A\ar[r]^{\eta_A}\ar[d] &
    \sharp A\ar[d]\\
    X\ar[r]^{\eta_X} &
    \sharp X
  }}\]
is a pullback.
(This says that $A$ has the ``initial cohesive structure'' induced from $X$: elements of $A$ cohere in precisely the ways that their images in $X$ cohere.)
For general $A\in\H/X$, we define $\sharpsub{X}A$ to be the pullback of $\sharp A \to \sharp X$ along $X\to \sharp X$.
When $\sharp$ preserves finite limits, this defines a reflective subfibration.

Reflective subfibrations constructed in this way are characterized by two special properties.
The first is that the relatively codiscrete morphisms are closed under composition.
A reflective subfibration with this property is equivalent to a \emph{stable factorization system}: a pair of classes of morphisms $(\E,\M)$ such that every morphism factors essentially uniquely as an \E-morphism followed by an \M-morphism, stably under pullback.
(The corresponding reflective subcategory of $\H/X$
% corresponding to such an $(\E,\M)$
is the category of \M-morphisms into $X$.)
%, with the factorization giving the reflector.)
If a reflective subcategory underlies a stable factorization system, its reflector preserve all pullbacks over objects in the subcategory --- and again, the converse holds in good situations~\cite{ShulmanSubfibrations}; such a reflector has \emph{stable units}~\cite{CHK}.

%We can express these two properties in terms of the internal type theory as follows.
% The first can be phrased as an axiom which appears to say that codiscrete objects are closed under dependent sums:
% \begin{equation}
%   \jd{\mathsf{ac}:\mathsf{isCodisc}(A), \mathsf{bc}:\textstyle\prod_{x:A}\mathsf{isCodisc}(B(x)) |- \mathsf{sc}: \mathsf{isCodisc}\left(\textstyle\sum_{x:A} B(x)\right)}\label{eq:depsum}
% \end{equation}
% But an equivalent, and often more convenient, phrasing is as follows.
We can axiomatize this property as follows.
Axiom~\eqref{eq:tsr} implies, in particular, a factorization operation:
\begin{gather}
  \jd{\mathsf{bc}:\mathsf{isCodisc}(B), f:A\to B |- \factsharp(f):\sharp A \to B} \label{eq:sharpfact1} \\
  \jd{\mathsf{bc}:\mathsf{isCodisc}(B), f:A\to B |- \mathsf{ff}_f:(\factsharp(f) \circ \eta_A = f)}\label{eq:sharpfact2}
\end{gather}
It turns out that relatively codiscrete morphisms are closed under composition if and only if we have a more general factorization operation, where $B$ may depend on $\sharp A$:
\begin{gather}
  \jd{ \mathsf{bc}:\textstyle\prod_{x:\sharp A}\mathsf{isCodisc}(B(x)),
    f:\textstyle\prod_{x:A} B(\eta_A(x))
    |- \factsharp(f):\textstyle\prod_{x:\sharp A} B(x)} \\
  \jd{ \mathsf{bc}:\textstyle\prod_{x:\sharp A}\mathsf{isCodisc}(B(x)),
    f:\textstyle\prod_{x:A} B(\eta_A(x))
    |- \mathsf{ff}_f: (\factsharp(f) \circ \eta_A = f)}
\end{gather}
This sort of ``dependent factorization'' is familiar in type theory;
it is related to~\eqref{eq:sharpfact1}--\eqref{eq:sharpfact2} in the same way that proof by induction is related to definition by recursion.

The second property is that if $g\in\E$ and $g f \in\E$, then $f\in \E$.
If a stable factorization system has this property, then it is determined by its underlying reflective subcategory, whose reflector must preserve finite limits.
(\E is the class of morphisms inverted by the reflector, and \M is defined by pullback as above.)
%we did for the relatively codiscrete morphisms above.)
We can state this in type theory as the preservation of $\sharp$-contractibility by homotopy fibers:
%, but here it is:
\begin{multline}\label{eq:lexrefl}
  \jd{\mathsf{acs}:\mathsf{isContr}(\sharp A), \mathsf{bcs}:\mathsf{isContr}(\sharp B),
    f:A\to B, b:B |- \mathsf{fcs} : \mathsf{isContr}\left(\sharp \textstyle\sum_{x:A}(f(x)=b)\right)}
\end{multline}
This completes our axiomatization of the reflective subcategory of codiscrete objects.
%, which externally means the adjunction $\Gamma \dashv \nabla$.
%As remarked above, nothing ensures that this subcategory is actually \ig, although since it is a left-exact-reflective subcategory of an \io-topos, it is itself an \io-topos.
%
We can apply the same reasoning to the reflective subcategory of \emph{discrete} objects.
%, i.e.\ to the adjunction $\Pi \dashv \Delta$.
%Instead of $\sharp$, in this case we abusively write $\Pi$ also for the reflector (which is technically the composite $\Delta\Pi$).
Now $\Pi$ does not preserve all finite limits, only finite products, so we cannot push the characterization all the way as we did for $\sharp$.
But because the target of $\Pi$ is \ig, it automatically has stable units; thus the discrete objects underlie \emph{some} stable factorization system $(\E,\M)$, which can be axiomatized as above with~\eqref{eq:lexrefl} omitted.
(For cohesion over a general base, we ought to demand stable units explicitly.)
% , regarding it as the ``indexed'' version of preserving finite products.)
%This gives another modality $\Pi$, which is a bit less well-behaved than $\sharp$.
%
We do not know whether there is a particular choice of such an $(\E,\M)$ to be preferred.
In \S\ref{sec:axiomatic-cohesion-ii}, we will mention a different way to axiomatize $\Pi$ which is less convenient, but does not require choosing $(\E,\M)$.

\subsection{Axiomatic cohesion II: $\sharp\type$}
\label{sec:axiomatic-cohesion-ii}

%edit
We consider now an axiomatization of the external hom $\H(A,B)$ in addition
to the internal hom $A \to B$. The latter we will equivalently write also as
$[A,B]$, which better matches the convention in algebraic topology and in geometry.
More precisely, we consider the external $\infty$-groupoid $\H(A,B)$, re-internalized as a codiscrete object.
(Since $\sharp$ preserves more limits than $\Pi$, the codiscrete objects are a better choice for this sort of thing.)

To construct such an external function-type, consider $\sharp\type$.
This is a codiscrete object, which as an $\infty$-groupoid is interpreted by the core of (some small full subcategory of) \H.
Any type $A:\type$ has an ``externalized'' version $\eta_\type(A):\sharp \type$, which we denote $\llbracket A\rrbracket$.
And since $\sharp$ preserves products, the operation
\begin{gather*}
[-,-]:\type\times\type\to\type \mathrlap{\qquad\qquad\text{induces an operation}}\\
[-,-]^\sharp:\sharp\type\times\sharp\type\to\sharp\type
\end{gather*}
which is interpreted by the internal-hom $[-,-]$ as an operation $\mathrm{Core}(\H) \times \mathrm{Core}(\H) \to \mathrm{Core}(\H)$.
We now define the \emph{escaping} morphism $\esc:\sharp\type\to\type$ as follows.%
%edit
\footnote{Technically, $\esc A$ lives in a higher universe than $A$, but we ignore this as it causes no problems.}
\[ \esc A \coloneqq \sum_{B:\sharp \sum_{X:\type} X} (\sharp(\mathsf{pr}_1)(B) = A) \]
Here $\sum_{X:\type} X$ is the type-theoretic version of the domain ${\type}_\bullet$ of the morphism
$p:{\type}_\bullet\to\type$ from \S\ref{sec:hott}, with $p$ being the first projection $\mathsf{pr}_1: \sum_{X:\type} X \to \type$.
Thus functoriality of $\sharp$ gives $\sharp(\mathsf{pr}_1):\sharp \sum_{X:\type} X \to \sharp \type$, and so $\sharp(\mathsf{pr}_1)(B) : \sharp \type$ can be compared with $A$.
%edit
Now it turns out that $\esc A$ is codiscrete (i.e.\ $\mathsf{isCodisc}(\esc A)$ is inhabited), and the composite
\[\type \xrightarrow{\eta_\type} \sharp\type \xrightarrow{\esc} \type\]
is equivalent to $\sharp$.
%Thus it makes sense to think of $\esc A$ as ``$\sharp A$'' for any $A:\sharp\type$.
Thus, since
in the intended categorical semantics
$\H(A,B) = \Gamma[A,B]$,
% and so $\nabla(\H(A,B)) = \sharp[A,B]$,
we can define the external function-type as
\[\H(A,B) \coloneqq \esc\left([A, B]^\sharp\right)\]
Note that this makes sense for any $A,B:\sharp\type$.
% Note that $A\toesc B$, unlike $A\tosharp B$, is a \emph{type}.
If instead $A,B:\type$, then $\H(\llbracket A\rrbracket, \llbracket B\rrbracket) \simeq \sharp[A,B]$.
Thus, if $f:[A,B]$, then applying this equivalence to $\eta_{[A,B]}(f): \sharp[A,B]$, we obtain an ``externalized'' version of $f$, which we denote $\llbracket f \rrbracket : \llbracket A\rrbracket \to \llbracket B\rrbracket$.
Now using dependent factorization for the modality $\sharp$, we can define all sorts of categorical operations externally.
We have composition:
\begin{equation*}
  \jd{A\cm B \cm C:\sharp\type, f:{\H(A, B)}, g:{\H(B, C)} |- g \circ f : \H(A,C)}
\end{equation*}
and the property of being an equivalence:
\[ \jd{A\cm B:\sharp \type, f:{\H(A,B)} |- \mathsf{eisEquiv}(f) : \type}. \]

Using these external tools, we can now complete our internal axiomatization of cohesion.
One may be tempted to define the coreflection $\flat$ as we did the reflections $\sharp$ and $\Pi$, but
this would amount to asking for a pullback-stable system of \emph{coreflective} subcategories of each $\H/X$, and at present we do not know any way to obtain this in models.
Instead, we work externally:
%, without implying anything about slice categories of \H.
\begin{gather}
  \mathsf{eisDisc} : [\sharp\type,\type]\\
  \jd{A:\sharp\type |- \mathsf{eisDiscIsProp}_A : \mathsf{isProp}(\mathsf{eisDisc}(A))}\\
  \flat:[\sharp\type,\sharp\type]\\
  \jd{A:\sharp\type |- \mathsf{flatIsDisc}_A : \mathsf{eisDisc}(\flat A)}\\
  \jd{A:\sharp \type |- \epsilon_A : \H(\flat A, A)}\label{flatCounit}\\
%which induces an equivalence $\H(A,\flat B) \simeq \H(A,B)$ by composition, if $A$ is discrete:
  \jd{A\cm B:\sharp \type, \mathsf{ad} : \mathsf{eisDisc}(A) |-
    \mathsf{flr} : \mathsf{isEquiv}(\lambda f^{{\H(A,\flat B)}} . \, \epsilon_B \circ f)}.\label{eq:flatUniv}
\end{gather}
%Here $\mathsf{eisDisc}:\sharp\type\to\type$ is a predicate expressing discreteness of a term in $\sharp\type$.
If we axiomatize discrete objects
% with a stable factorization system,
as in \S\ref{sec:axiomatic-cohesion-i}, we can define
$\mathsf{eisDisc}(A) \coloneqq \esc(\sharp(\mathsf{isDisc})(A))$.
% where $\mathsf{isDisc}:\type\to\type$ is the analogue for discrete objects of $\mathsf{isCodisc}$, to which we apply the functor $\sharp$ to get a function that can be applied to $A$, then finally escape the result to obtain a (codiscrete) type.
But we can also treat discrete objects externally only, with $\mathsf{eisDisc}$ axiomatic and $\Pi$ defined analogously to $\flat$.
% as a codiscrete predicate on $\sharp\type$, and $\Pi$ as a term $\sharp\type\to\sharp\type$ with dual properties to $\flat$.
This would allow us %, when interpreting the type theory in a cohesive \io-topos,
to avoid choosing an $(\E,\M)$ for the categorical interpretation.
In any case,~\eqref{eq:flatUniv} implies factorizations:
\begin{gather*}
  \jd{\mathsf{ad}:\mathsf{eisDisc}(A), f:{\H(A,B)} |- \factflat(f):{\H(A,\flat B)}} \\
  \jd{\mathsf{ad}:\mathsf{eisDisc}(A), f:{\H(A,B)} |- \mathsf{ff}_f:(\epsilon_B \circ \factflat(f) = f)}.
\end{gather*}
Thus, we can state the final axioms internally as
%that for any $A$, the induced maps $\flat A \to \flat\sharp A$ and $\sharp \flat A\to \sharp A$ are equivalences, can be stated internally as
\begin{gather}
  \jd{A:\sharp\type |- \mathsf{sfe}:\mathsf{eisEquiv} (\factsharp (\llbracket\eta_A\rrbracket \circ \epsilon_A))} \\
  \jd{A:\sharp\type |- \mathsf{fse}:\mathsf{eisEquiv} (\factflat (\llbracket\eta_A\rrbracket \circ \epsilon_A))}
\end{gather}
This completes the axiomatization of the internal homotopy type theory of a cohesive \io-topos, yielding the formal system that we call \emph{cohesive homotopy type theory}.

%%%%%%%%%%%%%%%%%%%%%%%%%%%%%%%%%%%%%%%%%%%%%%%%%%%%%%%%%%%%%%%%
\section{Quantum gauge field theory}
\label{sec:quantum}
%%%%%%%%%%%%%%%%%%%%%%%%%%%%%%%%%%%%%%%%%%%%%%%%%%%%%%%%%%%%%%%%

%With the axioms of \emph{cohesive homotopy type theory} in hand,
We now
give a list of constructions in this axiomatics
whose interpretation in cohesive
$(\infty,1)$-toposes \H reproduces various notions in differential geometry,
differential cohomology, geometric quantization and quantum gauge field theory.
Because of the homotopy theory built into the type theory, this also automatically generalizes
all these notions to homotopy theory.  For instance, a
\emph{gauge group} in the following may be interpreted as an ordinary gauge group such
as the Spin-group, but may also be interpreted as a \emph{higher} gauge
group, such as the String-2-group or the Fivebrane-6-group \cite{SSS}.
Similarly, all fiber products are automatically homotopy fiber products, and so on.
(The fiber product of $f:[A,C]$ and $g:[B,C]$ is $A\times_C B \coloneqq \sum_{x:A}\sum_{y:B}(f(x)=g(y))$, and this can be externalized easily.)

In this section we will mostly speak ``externally'' about a cohesive \io-topos \H.
This can all be expressed in type theory using the technology of \S\ref{sec:axiomatic-cohesion-ii}, but due to space constraints we will not do so.

%%%%%%%%%%%%%%%%%%%%%%%%%%%%%%%%%%%%%%%%%%%%%%%%%%%%%%%%%%%%%%%%
\subsection{Gauge fields}
%%%%%%%%%%%%%%%%%%%%%%%%%%%%%%%%%%%%%%%%%%%%%%%%%%%%%%%%%%%%%%%%

The concept of a \emph{gauge field} is usefully decomposed into two stages, the \emph{kinematical}
aspect and the \emph{dynamical} aspect bulding on that:

\hspace{-.6cm}
\begin{tabular}{|r||c|c|}
  \hline
  {\bf gauge field:} & {\bf kinematics} & {\bf dynamics}
  \\
  \hline
  \hline
  physics  term: &
  instanton sector / charge sector
  & gauge potential
  \\
  \hline
  formalized as: & cocycle in (twisted) cohomology & cocycle in (twisted) differential cohomology
  \\
  \hline
     diff.\ geo.\
     term:
  &
  fiber bundle
  &
  connection
  \\
  \hline
%  required
  ambient logic:
  & homotopy type theory & cohesive homotopy type theory
  \\
  \hline
\end{tabular}

%%%%%%%%%%%%%%%%%%%%%%%%%%%%%%%%%%%%%%%%%%%%%%%%%%%%%%%%%%%%%%%%
\subsubsection{Kinematics}
%%%%%%%%%%%%%%%%%%%%%%%%%%%%%%%%%%%%%%%%%%%%%%%%%%%%%%%%%%%%%%%%

Suppose $A$ is a pointed connected type, i.e.\ we have $a_0:A$ and $\pi_0(A)$ is contractible.
Then its loop type $\Omega A \coloneqq * \times_A * \simeq (a_0 = a_0)$ is
a group. This establishes an equivalence between pointed connected homotopy types and
group homotopy types; its inverse is called
\emph{delooping} and denoted $G \mapsto \mathbf{B}G$.\footnote{Currently, we cannot fully formalize completely general $\infty$-groups and their deloopings, because they involve infinitely many higher homotopies.  This is a mere technical obstruction that will hopefully soon be overcome.
It is not really a problem for us, since we generally care more about deloopings than groups themselves, and pointed connected types are easy to formalize.}
For instance, the
\emph{automorphism group} $\mathbf{Aut}(V)$ of a homotopy type $V$ is the looping of the image factorization
\[\xymatrix{
    {*} \ar@{->>}[r] \ar@/_1pc/[rr]|{\;\vdash V\;} & \mathbf{B}\mathrm{Aut}(V) \ar@{^{(}->}[r] & \type
  }.\]
It can also be defined as the space of self-equivalences, $\mathbf{Aut}(V) = \sum_{f:[V,V]}\mathsf{isEquiv}(f)$; this is roughly the content of the univalence axiom.

Given a group $G$ and a homotopy type $X$, we write
$
H^1(X,G) \coloneqq %\pi_0 \mathbf{H}(X, \mathbf{B}G) =
\pi_0 {\H(X,\mathbf{B}G)}
$
for the \emph{degree-1 cohomology} of $X$ with coefficients in $G$. If $G$ has higher
deloopings $\mathbf{B}^n G$, we write
$
H^n(X, G) \coloneqq % \pi_0 \mathbf{H}(X, \mathbf{B}^n G)
\pi_0 {\H(X,\mathbf{B}^n G)}
$
and speak of the \emph{degree-$n$ cohomology} of $X$ with coefficients in $G$.
The interpretation of this simple definition in homotopy type theory is very general,
and (if we allow disconnected choices of deloopings) much more general than what is traditionally called \emph{generalized cohomology}:
in traditional terms, it would be called \emph{non-abelian equivariant twisted sheaf hyper-cohomology}.

A  \emph{$G$-principal bundle} over $X$ is a function $p : P \to X$ where $P$ is
equipped with a $G$-action over $X$ and such that $p$ is the quotient $P \to P\ssl G$
(as always, this is a \emph{homotopy quotient}, constructible as a higher inductive type
%edit:
~\cite{ShulmanLumsdaine,hottbook},%ShulmanHITs,LumsdaineHITs,
, see above.
One finds that the delooping $\mathbf{B}G$
is the \emph{moduli stack} of $G$-principal bundles: for any $g : X \to \mathbf{B}G$, its
(homotopy) fiber is canonically a $G$-principal bundle over $X$, and this establishes
an equivalence
$G \mathrm{Bund}(X) \simeq {\H(X,\mathbf{B}G)} $
%\mathbf{H}(X,\mathbf{B}G) \simeq \Gamma [X, \mathbf{B}G]$
between $G$-principal bundles and cocycles in \emph{$G$-cohomology}.
In particular, equivalence classes of $G$-principal bundles on $X$ are classified by $H^1(X,G)$.

Conversely, any $G$-action $\rho : G \times V \to V$ is equivalently encoded in
a fiber sequence
$
  \xymatrix@-1pc{
    V \ar[r] & V\ssl G \ar[r]^{\bar \rho}
  & \mathbf{B}G
  }
$,
hence in a $V$-fiber bundle $V\ssl G$ over $\mathbf{B}G$.
This is the \emph{universal $\rho$-associated} $V$-fiber bundle
in that, for $g$ and $P$ as above, the $V$-bundle $E \coloneqq P \times_G V \to X$
is equivalent to the pullback $g^* \bar \rho$.

%new paragraph
Syntactically, this means that homotopy type theory in the context of a pointed connected type
$\mathbf{B}G$ is the \emph{representation theory} of $G$: the fiber sequence above
is the interpretation of a dependent type
$
  \jd{x : \mathbf{B}G |- V : \type}
$
which hence encodes a $G$-$\infty$-representation on a type $V$.
One finds that the dependent product
$
 \vdash( \prod_{x : \mathbf{B}G} V : \type)
$
is the type of \emph{invariants} of the $G$-action, while the dependent sum
$
  \vdash(\sum_{x : \mathbf{B}G} V : \type)
$
is the syntax for the quotient $V\sslash G$ mentioned above. Moreover, for $V_1, V_2$ two
such $G$-representations, the dependent product of their function type
$
  \vdash (\prod_{x : \mathbf{B}G} [V_1, V_2] : \type)
$
is interpreted as the type of $G$-action homomorphisms, while the dependent sum
$
  \vdash (\sum_{x : \mathbf{B}G} [V_1, V_2] : \type)
$
is interpreted
%edit
as
the quotient $[V_1, V_2]\sslash G$ of the space of all morphisms modulo
the \emph{conjugation action} by $G$. It follows that the homotopy type of sections
$\Gamma_X(E)$ above is the interpretation of the syntax
$\vdash (\prod_{x : \mathbf{B}G}  [P, V] : \type)$. Notice that this
expression reproduces almost verbatim the traditional statement that a section of a $\rho$-associated
$V$-bundle is a $G$-equivariant map from the total space $P$ to $V$, only that the
interpretation of this statement in homotopy type theory here generalizes it to higher bundles.

In particular, the
homotopy type of sections $\Gamma_X(E)$
of the associated bundle $E$ above is equivalent to the dependent product of the mapping
space $[g, \bar \rho]$ formed in the slice, the
homotopy type theory syntax for it being $\vdash( \prod_{x : \mathbf{B}G} [P, V] : \type)$.

Since all the bundles involved are locally trivial with respect to the intrinsic notion of
covers (epimorphic maps) it follows that
elements of $\Gamma_X(E)$ are locally maps to $V$. If $V$ here is pointed connected,
and hence $V \simeq \mathbf{B}H$, then $E$ is called an \emph{$H$-gerbe} over $X$.
In this case a section of $E \to X$ is therefore locally a cocycle in $H$-cohomology,
and hence globally a cocycle in \emph{$g$-twisted $H$-cohomology} with respect to the
\emph{local coefficient bundle} $E \to X$. Hence twisted cohomology in $\mathbf{H}$
is ordinary cohomology in a slice $\mathbf{H}_{/\mathbf{B}G}$. All this is discussed in
detail in \cite{NSS}.

In gauge field theory a group $H$ as above serves as the \emph{gauge group} and then
an $H$-principal bundle on $X$ is the charge/kinematic part of an \emph{$H$-gauge field} on $X$.
(In the special case that $H$ is a discrete homotopy type, this is already the full
gauge field, as in this case the dynamical part is trivial).
The mapping type $[X, \mathbf{B}H]$ interprets as the moduli stack of kinematic
$H$-gauge fields on $X$, and a term of identity type $\lambda :  (\phi_1 = \phi_2)$
is a \emph{gauge transformation} between two gauge field configurations $\phi_1, \phi_2: [X, \mathbf{B}H]$.

It frequently happens that the charge of one $G$-gauge field $\Phi$ \emph{shifts} another $H$-gauge field
$\phi$, in generalization of the way that magnetic charge shifts the electromagnetic field.
Such shifts are controlled by an action $\rho$ of $G$ on $\mathbf{B}H$ and in this case
$\Phi$ is a cocycle in $G$-cohomology and $\phi$ is a cocycle in $\Phi$-twisted $H$-cohomology
with respect to the local coefficient bundle $\bar \rho$.

%new paragraph
An example of this phenomenon is given by the field configurations in Einstein-gravity
(general relativity). For this we assume that the \emph{general linear group} $\mathrm{GL}(n)$
in dimension $n$ exists as a group type, which is the case in the standard models for
smooth cohesion. Notably if we pass to a context of synthetic differential cohesion then
the first order infintesimal neighbourhood of the origin of $\mathbb{R}^n$ exists as a
type $D^n$ and we have $\mathrm{GL}(n) = \mathbf{Aut}(D^n)$. By the above, for an $n$-dimensional
smooth manifold $\Sigma$ (to be thought of as \emph{spacetime}) the frame bundle
$T \Sigma \to \Sigma$
%edit
is characterized as the type sitting
in a fiber sequence
$$
  \raisebox{20pt}{
  \xymatrix{
    T \Sigma \ar[r] & \Sigma \ar[d]
  \\
  & \mathbf{B}GL(n)
  }}
  \,,
$$
which syntactically is therefore the dependent type $\jd{x : \mathbf{B}GL(n) |- T \Sigma : \type}$.
Let then $\mathbf{B}O(n) \to \mathbf{B}\mathrm{GL}(n)$ be the delooping of the inclusion of the
maximal compact subgroup
%edit
\footnote{Assuming that we can talk about general colimits, which is
work in progress in fully formal HoTT, then geometric compactness can be axiomatized
as smallness with respect to filtered colimits whose structure maps are monomorphisms.}
, the
orthogonal group, into the general linear group. This sits in a homotopy fiber sequence
$$
  \raisebox{20pt}{
  \xymatrix{
    \mathrm{GL}(n)/O(n) \ar[r] & \mathbf{B}O(n) \ar[d]
  \\
  & \mathbf{B}\mathrm{GL}(n)
  }
  }
  \,,
$$
with the smooth coset space, exhibiting the canonical $\mathrm{GL}(n)$-action on that space.
Syntactically this is therefore the dependent type
$\jd{x : \mathbf{B}\mathrm{GL}(n) |- \mathrm{GL}(n)/O(n) : \type}$.
With the above one finds then that the homotopy type theory syntax
$$
  \vdash\; \left(\prod_{x : \mathbf{B}\mathrm{GL}(n)} [T \Sigma, \mathrm{GL}(n)/O(n)] : \type\right)
$$
is interpreted as the space of diagrams
$$
  \Gamma_{\Sigma}(T \Sigma \times_{GL(n)} \mathrm{GL}(n)/O(n))
  =
  \left\{
   \raisebox{20pt}{
  \xymatrix{
    \Sigma \ar[rr]_{\ }="s" \ar[dr]^{\ }="t" && \mathbf{B}O(n) \ar[dl]
  \\
  & \mathbf{B}\mathrm{GL}(n)
  \ar@{=>} "s"; "t"
  }}
  \right\}
  \,.
$$
Here $T \Sigma \times_{\mathrm{GL}(n)} \mathrm{GL}(n)/O(n)$ denotes the $(\mathrm{GL}(n)/O(n))$-bundle associated to the frame bundle (so $\times_{\mathrm{GL}(n)}$ denotes not a pullback but a tensor product over $\mathrm{GL}(n)$-actions).
The homotopy type theory syntax here is manifestly the expression for $\mathrm{GL}(n)$-equivariant maps from the
frame bundle to the coset, while the diagram manifestly expresses a reduction of the
structure group of the frame bundle to an $O(n)$-bundle. Both of these equivalent
constructions encode a choice of \emph{vielbein}, hence of a \emph{Riemannian metric}
on $\Sigma$, hence of a field of gravity. (We could just as well discuss
genuine pseudo-Riemannian metrics.) Moreover, by the homotopy-theoretic construction
this space of fields automatically contains the $O(n)$-gauge transformations on the vielbein
fields corresponding to choices of \emph{reference frames}.
But since gravity is a \emph{generally covariant} field theory,  the correct
configuration space of gravity is furthermore the \emph{quotient} of this space of
vielbein fields by the diffeomorphism action on $T \Sigma$. One finds from the above that
in homotopy type theory syntax this is simply expressed by interpreting the
space of fields in the context of the delooped automorphism group of $\Sigma$
and then forming the dependent sum:
the interpretation of
$$
  \vdash\;\left( \prod_{x : \mathbf{B}\mathrm{GL}(n)} \;\sum_{y : \mathbf{B}\mathbf{Aut}(T \Sigma)}
    [T \Sigma, \mathrm{GL}(n)/O(n)] : \type\right)
$$
is the moduli stack of the generally covariant field of gravity. In the
smooth context it is the Lie integration of the gravitational (off-shell) BRST complex with diffeomorphism
ghosts in degree 1.

Further discussion of examples and further
pointers are in \cite{SchreiberLectures, SchreiberErlangen}.

%%%%%%%%%%%%%%%%%%%%%%%%%%%%%%%%%%%%%%%%%%%%%%%%%%%%%%%%%%%%%%%%
\subsubsection{Dynamics}
\label{Dynamics}
%%%%%%%%%%%%%%%%%%%%%%%%%%%%%%%%%%%%%%%%%%%%%%%%%%%%%%%%%%%%%%%%

Given a $G$-principal bundle in the presence of cohesion, we may ask if its cocycle $g : X \to \mathbf{B}G$
lifts through the counit $\epsilon_{\mathbf{B}G} : \flat \mathbf{B}G \to \mathbf{B}G$
from (\ref{flatCounit}) to a cocycle
$\nabla : X \to \flat \mathbf{B}G$. By the $(\Pi \dashv \flat)$-adjunction this is equivalently
a map $\Pi(X) \to \mathbf{B}G$. Since $\Pi(X)$ is interpreted as the \emph{path $\infty$-groupoid}
of $X$, such a $\nabla$ is a \emph{flat parallel transport} on $X$ with values in $G$, equivalently
a \emph{flat $G$-principal connection} on $X$.

Consider then the homotopy fiber
$
  \flat_{\mathrm{dR}} \mathbf{B}G \coloneqq \flat \mathbf{B}G \times_{\mathbf{B}G} {*}
$.
By definition, %this homotopy type is such that
a map $\omega : X \to \flat_{\mathrm{dR}} \mathbf{B}G$ is a
flat $G$-connection on $X$ together with a trivialization of the underlying
$G$-principal bundle. This is interpreted
as a \emph{flat $\mathfrak{g}$-valued differential form}, where $\mathfrak{g}$ is the \emph{Lie algebra}
of $G$. By using this definition in the statement of the above classification of $G$-principal bundles,
one finds that every flat connection
$\nabla : X \to \flat \mathbf{B}G$ is \emph{locally} given by a flat $\mathfrak{g}$-valued form:
$\nabla$ is equivalently a form $A : P \to \flat_{\mathrm{dR}} \mathbf{B}G$ on the total space
of the underlying $G$-principal bundle, such that this is $G$-equivariant in a natural sense.
Such an $A$ is interpreted as the incarnation of the connection $\nabla$
in the form of an \emph{Ehresmann connection} on $P \to X$.

Moreover, the coefficient $\flat_{\mathrm{dR}}\mathbf{B}G$ sits in
a long fiber sequence of the form
\[
  \xymatrix{
    G
   \ar[r]^-{\theta_G}
    &
    \flat_{\mathrm{dR}}\mathbf{B}G
   \ar[r]
  &
    \flat \mathbf{B}G
   \ar[r]^-{\epsilon_{\mathbf{B}G}}
  &
  \mathbf{B}G
  }
  \,.
\]
with the further homotopy fiber $\theta_{G}$
giving a canonical flat $\mathfrak{g}$-valued differential form on $G$. This is
the \emph{Maurer-Cartan form} of $G$, in that when interpreted in smooth homotopy types
and for $G$ an ordinary Lie group, it is canonically identified with the classical
differential-geometric object of this name. Here in cohesive homotopy type theory it
exists in much greater generality.

Specifically, assume that $G$ itself is once more
deloopable, hence assume that $\mathbf{B}^2 G$ exists. Then
the above long fiber sequence extends further to the right as
$
  \xymatrix{
    \flat \mathbf{B}G
   \ar[r]^-{\epsilon_{\mathbf{B}G}}
  &
  \mathbf{B}G
   \ar[r]^-{\theta_{\mathbf{B}G}}
  &
  \flat_{\mathrm{dR}} \mathbf{B}^2 G
  }
$, since $\flat$ is right adjoint.
This means, by the universal property of homotopy fibers, that if $g : X \to \mathbf{B}G$ is the
cocycle for a $G$-principal bundle on $X$, then the class of the differential form $\theta_{\mathbf{B}G}(g)$ is the \emph{obstruction} to the existence of a flat
connection $\nabla$ on this bundle. Hence
this class is interpreted as the \emph{curvature} of the bundle, and we interpret
the Maurer-Cartan form
$\theta_{\mathbf{B}G}$ of the delooped group $\mathbf{B}G$
as the \emph{universal curvature characteristic} for $G$-principal bundles.

This universal curvature characteristic is the key to the notion of non-flat connections,
for it allows us to define these in the sense of twisted cohomology as
\emph{curvature-twisted flat cohomology}.
There is, however, a choice involved in defining the universal curvature-twist, which depends
on the intended application. But in standard interpretations there is a collection
of types singled out, called the \emph{manifolds}, and the standard universal
curvature twist can then be characterized as a map
$i : \Omega^{2}_{\mathrm{cl}}(-,\mathfrak{g}) \to \flat_{\mathrm{dR}}\mathbf{B}^2 G$ out of a 0-truncated
homotopy type such that for all
manifolds $\Sigma$ its image under $[\Sigma,-]$ is epi, meaning that
$\Omega^2_{\mathrm{cl}}(\Sigma,\mathfrak{g}) \to [\Sigma, \flat_{\mathrm{dR}}\mathbf{B}^2 G]$
is an \emph{atlas} in the sense of geometric stack theory.

Assuming such a choice of universal curvature twists has been made, we may then define
the moduli
of general (non-flat) $G$-principal connections to be the homotopy fiber product
\[
  \mathbf{B}G_{\mathrm{conn}}
  \coloneqq
  i^* \theta_{\mathbf{B}G}
  =
  \mathbf{B}G \times_{\flat_{\mathrm{dR}} \mathbf{B}^2 G} \Omega^2(-,\mathfrak{g})
  \,.
\]
In practice one is usually interested in a canonical abelian (meaning arbitrarily deloopable, i.e.\ $E_\infty$)
group $A$ and the tower of delooping groups $\mathbf{B}^n A$ that it induces.
In this case we write
$
  \mathbf{B}^n A_{\mathrm{conn}}
  \coloneqq
  \mathbf{B}^n A \times_{\flat_{\mathrm{dR}} \mathbf{B}^{n+1} A} \Omega^{n+1}(-,A)
$.
When interpreted in smooth homotopy types and choosing the Lie group
$A = \mathbb{C}^\times$ or $= U(1)$ one finds that $\mathbf{B}^n A_{\mathrm{conn}}$
is the coefficient for \emph{ordinary differential cohomology}, specifically that it is presented
by the $\infty$-stack given by the \emph{Deligne complex}.

Notice that the pasting law for homotopy pullbacks implies generally that the restriction of
$\mathbf{B} G_{\mathrm{conn}}$ to
vanishing curvature indeed coincides with the universal flat coefficients:
$\flat \mathbf{B}G \simeq \mathbf{B}G_{\mathrm{conn}} \times_{\Omega^2(-,\mathfrak{g})} \{*\}$.
This means that we obtain a factorization
of $\epsilon_{\mathbf{B} G}$ as $\flat \mathbf{B}G \to \mathbf{B}G_{\mathrm{conn}} \to \mathbf{B}G$.

Let then $G$ be a group which is not twice deloopable, hence to which the above universal definition
of $\mathbf{B}G_{\mathrm{conn}}$ does not apply. If we have in addition a map
$\mathbf{c} : \mathbf{B}G \to \mathbf{B}^n A$ given (representing a
universal characteristic class in
$H^n(\mathbf{B}G, A)$), then we may still ask for \emph{some} homotopy type
$\mathbf{B}G_{\mathrm{conn}}$ that supports a \emph{differential refinement}
$
 \mathbf{c}_{\mathrm{conn}} : \mathbf{B}G_{\mathrm{conn}} \to \mathbf{B}^n A_{\mathrm{conn}}
$
of
$\mathbf{c}$
in that it lifts the factorization of $\epsilon_{\mathbf{B}^n A}$ by
$\mathbf{B}^n A_{\mathrm{conn}}$.
Such a $\mathbf{c}_{\mathrm{conn}}$ interprets, down on cohomology, as a
\emph{secondary universal characteristic class} in the sense of refined Chern-Weil theory.
Details on all this are in \cite{FSSt, SchreiberCohesion}.

With these choices and
for $G$ regarded as a gauge group, a genuine \emph{$G$-gauge field} on $\Sigma$ is a map
$\phi : \Sigma \to \mathbf{B}G_{\mathrm{conn}}$. For $G$ twice deloopable, the \emph{field strength}
of $\phi$ is the
composite
$F_{\phi} : \Sigma \to \mathbf{B}G_{\mathrm{conn}} \xrightarrow{F_{(-)}} \Omega^2_{\mathrm{cl}}(-,\mathfrak{g})$.

Moreover, the choice of $\mathbf{c}_{\mathrm{conn}}$ specifies an exended \emph{action functional}
on the moduli type $[\Sigma, \mathbf{B}G_{\mathrm{conn}}]$ of $G$-gauge field configurations, and hence
specifies an actual quantum gauge field theory. This we turn to now.

%%%%%%%%%%%%%%%%%%%%%%%%%%%%%%%%%%%%%%%%%%%%%%%%%%%%
\subsection{$\sigma$-Model QFTs}
%%%%%%%%%%%%%%%%%%%%%%%%%%%%%%%%%%%%%%%%%%%%%%%%%%%%%

An $n$-dimensional (``nonlinear'')
\emph{$\sigma$-model} quantum field theory describes the dynamics of an $(n-1)$-dimensional
quantum ``particle'' (for instance an electron for $n = 1$, a string  for $n = 2$, and generally
an ``$(n-1)$-brane'') that propagates in a \emph{target space} $X$
(for instance our spacetime) while acted on by forces (for instance the \emph{Lorentz force})
exerted by a fixed background $A$-gauge field on $X$
(for instance the electromagnetic field for $n = 1$ or the \emph{Kalb-Ramond $B$-field} for $n = 2$
or the \emph{supergravity $C$-field} for $n = 3$).
By the above, this background gauge field is the interpretation of a map
$
  \mathbf{c}_{\mathrm{conn}}
    : X \to  \mathbf{B}^n A_{\mathrm{conn}}
$.

Let then $\Sigma$ be a cohesive homotopy type of \emph{cohomological dimension} $n$,
to be thought of as the abstract \emph{worldvolume} of the $(n-1)$-brane.
The homotopy type $[\Sigma,X]$ of cohesive
maps from $\Sigma$ to $X$ is interpreted as the moduli space of field configurations of the
$\sigma$-model for this choice of shape of worldvolume.
The (gauge-coupling part of) the \emph{action functional} of the $\sigma$-model is
then to be the \emph{$n$-volume holonomy} of the background gauge field over a given
field configuration $\Sigma \to X$.

To formalize this, we need the notion of \emph{concreteness}.
If $X$ is a cohesive homotopy type, its \emph{concretization} is the image factorization
$X \twoheadrightarrow \mathrm{conc} X \hookrightarrow \sharp X$
% $
%   \xymatrix@-1pc{
%     X \ar@{->>}[r] & \mathrm{conc} X \ar@{^{(}->}[r] & \sharp X
%   }
% $
of $\eta_X:X\to \sharp X$ (\ref{SharpUnit}).
We call $X$ \emph{concrete} if $X \to \mathrm{conc}\,X$ is an equivalence.
In the standard smooth model, the 0-truncated concrete
cohesive homotopy types are precisely the \emph{diffeological spaces} (see~\cite{SmoothSpaces}).
Generally, for models over \emph{concrete sites}, they are the \emph{concrete sheaves}.

Now we can define the \emph{action functional} of the $\sigma$-model associated
to the background gauge field $\mathbf{c}_{\mathrm{conn}}$ to be the composite
$$
  \exp(i S(-))
    :
  \xymatrix{
    [\Sigma, X]
  \ar[rr]^-{[\Sigma, \mathbf{c}_{\mathrm{conn}}]}
  &&
  [\Sigma, \mathbf{B}^n A_{\mathrm{conn}}]
  \ar[r]
  &
  \mathrm{conc}\,\pi_0 [\Sigma, \mathbf{B}^n A_{\mathrm{conn}}]
  }
  \,.
$$
In the standard smooth model, with $A = \mathbb{C}^\times$ or
$U(1)$, the second morphism is \emph{fiber integration in differential cohomology}
$\exp(2 \pi i \int_{\Sigma}(-))$.
For $n = 1$ this computes the line holonomy of a circle bundle with connection,
hence the correct gauge coupling action functional of the 1-dimensional $\sigma$-model; for
$n = 2$ it computes the surface holonomy of a circle 2-bundle, hence the
correct ``WZW-term'' of the string; and so on.

Traditionally, $\sigma$-models are thought of as having as target space $X$ a manifold
or at most an orbifold. However, since these are smooth homotopy $n$-types for $n \leq 1$, it is natural to
allow $X$ to be a general cohesive homotopy type.
If we do so, then a variety of
quantum field theories that are not traditionally considered as $\sigma$-models become
special cases of the above general setup.
Notably, if $X = \mathbf{B}G_{\mathrm{conn}}$ is
the moduli for $G$-principal connections, then a $\sigma$-model with target space $X$ is
a $G$-gauge theory on $\Sigma$. Moreover, as we have seen above, in this case the background
gauge field is a secondary universal characteristic invariant. One finds that the corresponding
action functional
$
 \exp(2 \pi i \int_\Sigma [\Sigma, \mathbf{c}_{\mathrm{conn}}])
 : [\Sigma, \mathbf{B}G_{\mathrm{conn}}] \to A
$
is that of \emph{Chern-Simons}-type gauge field theories \cite{FSS, CS},
including the standard 3-dimensional Chern-Simons theories as well a
various higher generalizations.

More generally, at least in the smooth model, there is a \emph{transgression} map for
differential cocycles: for
$\Sigma_d$ a manifold of dimension $d$ there is a canonical map
$$
  \exp(2 \pi i \int_{\Sigma}[\Sigma,\mathbf{c}_{\mathrm{conn}}])
  :
  \xymatrix{
    [\Sigma_d, X]
  \ar[rr]^-{[\Sigma_d, \mathbf{c}_{\mathrm{conn}}]}
  &&
  [\Sigma_d, \mathbf{B}^n A_{\mathrm{conn}}]
  \ar[rr]^{\exp(2 \pi i \int_\Sigma)}
  &&
  \mathbf{B}^{n-d} A_{\mathrm{conn}}
  }
$$
modulating an $n$-bundle on the $\Sigma_d$-mapping space. For $d = n-1$ and
quadratic Chern-Simons-type theories this turns out to be the (off-shell)
prequantum bundle of the QFT. See \cite{FSS} for details on these matters.
Thus the differential characteristic
cocycle $\mathbf{c}_{\mathrm{conn}}$ should itself be regarded as a
\emph{higher prequantum bundle}, in the sense we now discuss.

%%%%%%%%%%%%%%%%%%%%%%%%%%%%%%%%%%%%%%%%%%%%%%%%%%%%%%%%%%%%%%%%
\subsection{Geometric quantization}
\label{GeometricQuantization}
%%%%%%%%%%%%%%%%%%%%%%%%%%%%%%%%%%%%%%%%%%%%%%%%%%%%%%%%%%%%%%%%

Action functionals as above are supposed to induce $n$-dimensional \emph{quantum} field theories
by a process called \emph{quantization}. One formalization of what this means is
\emph{geometric quantization}, which is well-suited to formalization in
cohesive homotopy type theory. We indicate here how to formalize the spaces of
higher (pre)quatum states that an extended QFT assigns in codimension $n$.

The \emph{critical locus} of a
local action functional -- its \emph{phase space} or
\emph{Euler-Lagrange solution space} -- carries a canonical closed 2-form $\omega$,
and standard geometric quantization gives a method for constructing the
\emph{space of quantum states} assigned by the QFT in dimension $n-1$
as a space of certain sections of a
\emph{prequantum bundle} whose curvature is $\omega$.
This works well for non-extended topological quantum field theories and generally for
$n = 1$ (quantum mechanics). The generalization to $n \geq 2$ is
called \emph{multisymplectic} or \emph{higher symplectic geometry} \cite{CR},
for here $\omega$ is promoted to an $(n+1)$-form which reproduces the former 2-form
upon transgression to a mapping space. Exposition of the string $\sigma$-model
($n = 2$) in the context of higher symplectic geometry is in \cite{BHR},
and discussion of quantum Yang-Mills theory ($n = 4$)
and further pointers are in \cite{Kan}.
A homotopy-theoretic formulation is given in \cite{SchreiberErlangen}: here the prequantum
bundle is promoted to a prequantum $n$-bundle, a $(\mathbf{B}^{n-1}A)$-principal connection
as formalized above.

Based on this we can give a formalization of central ingredients of geometric quantization in
cohesive homotopy type theory. When interpreted in the standard smooth model with $A = \mathbb{C}^\times$ or $= U(1)$ the following reproduces the traditional notions for $n = 1$, and for $n \geq 2$ consistently generalizes them to higher
geometric quantization.

Let $X$ be any cohesive homotopy type. A closed $(n+1)$-form on $X$ is a map
$\omega : X \to \Omega^{n+1}_{\mathrm{cl}}(-,A)$, as discussed in section \ref{Dynamics}.
We may call the pair $(X, \omega)$ a \emph{pre-$n$-plectic} cohesive homotopy type.
The group of \emph{symplectomorphisms} or \emph{canonical transformations} of $(X, \omega)$
is the automorphism group of $\omega$:
\[
    \mathbf{Aut}_{/\Omega^{n+1}_{\mathrm{\mathrm{cl}}}(-, A)}(\omega)
  =
  \left\{
    \raisebox{20pt}{
    \xymatrix{
       X \ar[rr]^{\simeq} \ar[dr]_{\omega} && X \ar[dl]^{\omega}
     \\
     & \Omega^{n+1}_{\mathrm{cl}}(-,A)
    }
    }
  \right\}
  \,,
\]
regarded as an object in the slice
$\mathbf{H}_{/\Omega^{n+1}_{\mathrm{cl}}(-,A)}$.
A \emph{prequantization} of $(X,\omega)$ is a lift
\[ %\hspace{-10pt}$
 \raisebox{18pt}{
  \xymatrix{
      & \mathbf{B}^n A_{\mathrm{conn}}
    \ar[d]^{F_{(-)}}
     \\
     X \ar[ur]^-{\mathbf{c}_{\mathrm{conn}}} \ar[r]^-\omega & \Omega^{n+1}_{\mathrm{cl}}(-,A)
  }
  }
\]
through
the defining projection from the moduli of $(\mathbf{B}^{n-1}A)$-principal connections.
This $\mathbf{c}_{\mathrm{conn}}$ modulates the \emph{prequantum $n$-bundle}.
Since $A$ is assumed abelian, there is abelian
group structure on $\pi_0(X \to \mathbf{B}^n A_{\mathrm{conn}})$ and hence we may rescale
$\mathbf{c}_{\mathrm{conn}}$ by a natural number $k$.
This corresponds to rescaling \emph{Planck's constant} $\hbar$ by $1/k$. The limit $k \to \infty$
in which $\hbar \to 0$ is the \emph{classical limit}.

The automorphism group of the prequantum bundle
\[
   \mathbf{Aut}_{/\mathbf{B}^n A_{\mathrm{conn}}}(\mathbf{c}_{\mathrm{conn}})
  :=
  \left\{
    \raisebox{20pt}{
    \xymatrix{
      X \ar[rr]^{\sigma}_{\ }="s" \ar[dr]_{\mathbf{c}_{\mathrm{conn}}}^{\ }="t" & & X \ar[dl]^{\mathbf{c}_{\mathrm{conn}}}
    \\
    & \mathbf{B}^n A_{\mathrm{conn}}
    \ar@{=>}^\simeq "s"; "t"
    }
    }
  \right\},
\]
in the slice $\mathbf{H}_{/\mathbf{B}^n A_{\mathrm{conn}}}$,
is the \emph{quantomorphism group} of the system.
%new paragraph
Syntactically this is
$$
  \vdash\; \left(\prod_{\nabla : \mathbf{B}^n A_{\mathrm{conn}}}
    \mathbf{Aut}(X(\nabla))
  :
  \type
  \right).
$$
See \cite{SchreiberErlangen} for more on this.

There is an evident projection from the
quantomorphism group to the symplectomorphism group, and its image is the group of
\emph{Hamiltonian symplectomorphisms}.
The Lie algebra of the quantomorphism group is that of
\emph{Hamiltonian observables} equipped with the \emph{Poisson bracket}.
If $X$ itself has abelian group structure,
then the subgroup of the quantomorphism group covering the action on $X$ on itself is the
\emph{Heisenberg group} of the system.
An action of any group $G$ on $X$ by quantomorphisms, i.e.\ a map
$\mu : \mathbf{B}G \to \mathbf{B}\mathbf{Aut}_{/\mathbf{B}^n A_{\mathrm{conn}}}(\mathbf{c}_{\mathrm{conn}})$,
is a \emph{Hamiltonian $G$-action} on $(X,\omega)$. The (homotopy) quotient
$\mathbf{c}_{\mathrm{conn}}\ssl G : X \ssl G \to \mathbf{B}^n A_{\mathrm{conn}}$ is the
corresponding \emph{gauge reduction} of the system.

After a choice of representation $\rho$ of $\mathbf{B}^{n-1}A$ on some $V$, the
space of \emph{prequantum states} is
\[
 \Gamma_X(E)
   \coloneqq
  [\mathbf{c}, \mathbf{p}]_{/\mathbf{B}^n A}
  =
  \left\{
    \raisebox{20pt}{
    \xymatrix{
      X \ar[rr]^{\sigma}_{\ }="s" \ar[dr]_{\mathbf{c}}^{\ }="t" & & V \ssl {\mathbf{B}^{n-1}A}
    \ar[dl]^{\bar \rho}
    \\
    & \mathbf{B}^n A
    \ar@{=>}^\simeq "s"; "t"
    }
    }
  \right\},
\]
the space of $\mathbf{c}$-twisted cocycles with respect to the local coefficient bundle
$\bar \rho$.
There is an evident action of the quantomorphism group on $\Gamma_X(E)$
and this is the action of \emph{prequantum operators} on the space of states.

It remains to formalize in cohesive homotopy type theory
the quantization step from this \emph{pre}quantum data to actual quantized
field theory. This is discussed, in terms of standard homotopy theory and
$\infty$-topos theory,  in \cite{NuitenThesis},
to which we refer the reader for further details.
This quantization step involves, naturally, linear algebra, which
internal to homotopy theory
is \emph{stable homotopy theory}, the theory of spectra and
ring spectra. This has not been fully formalized in
homotopy type theory at the moment, but it seems clear that this can be done.
In any case, a discussion of the quantization step
in the language of homotopy type theory is possible, but
beyond the scope of this article.

\medskip

In conclusion then, the amount of gauge QFT notions naturally
formalized here in cohesive homotopy type theory seems to be remarkable,
emphasizing the value of a formal, logical, approach to concepts like
smoothness and cohomology.

\bibliographystyle{eptcs}
\bibliography{cohesion-rev}

\end{document}